\documentclass[12p]{article}
\usepackage{graphicx} 
\usepackage[margin=1in]{geometry}
\usepackage{lscape}
\usepackage{pdflscape}
\usepackage{amsmath,amssymb}
\usepackage{tabularx}
\usepackage{booktabs}
\usepackage{bbm}
\usepackage{ccaption}
\usepackage{dcolumn}
\usepackage{multirow}
\usepackage{nicefrac}
\usepackage{float}
\usepackage{subcaption}
\usepackage{comment}
\usepackage{xcolor}
\usepackage[onehalfspacing,nodisplayskipstretch]{setspace}
\usepackage{color}
\usepackage{url}
\usepackage{varioref}
\usepackage{eurosym}
\usepackage{mdframed}
\usepackage{rotating}
\usepackage{enumitem}
\newtheorem{theorem}{Theorem}

\newtheorem{assumption}{Assumption}
\newtheorem{estimand}{Estimand}

\newtheorem{proposition}[theorem]{Proposition}

\newenvironment{proof}[1][Proof]{\noindent\textbf{#1.} }{\ \rule{0.5em}{0.5em}}
\usepackage{natbib}
\definecolor{navy}{rgb}{0,.3,.7}
\definecolor{myorange}{RGB}{238, 166, 63 }
\definecolor{myblue}{RGB}{29, 130, 231 }
\definecolor{green}{RGB}{49, 178, 54}
\definecolor{darkred}{RGB}{169,5,5}
\definecolor{darkgreen}{RGB}{0,152,0}
\usepackage[pdfpagemode=UseThumbs,%
     colorlinks=true,%
     urlcolor=navy,%
     citecolor=navy,%
     linkcolor=red]{hyperref}
\usepackage{cleveref}
\graphicspath{{figures/}}
\usepackage[font=bf]{caption} 
\newcolumntype{d}[1]{D{.}{.}{#1}}
\newcolumntype{e}{D{.}{.}{-1}}
\newcolumntype{.}{D{.}{.}{3}}
\newcommand{\E}{\mathbb{E}}
\newcommand{\bs}{\boldsymbol}

\begin{document}

%%%%%%%%%%%%%%%%%%%%%%%%%%%%%%%%%%%%%%%%%%%%%%%%%%%%%%%%%%%%%%%%%%%%
%  TITLE PAGE
%%%%%%%%%%%%%%%%%%%%%%%%%%%%%%%%%%%%%%%%%%%%%%%%%%%%%%%%%%%%%%%%%%%%

\title{\large Difference-in-Differences in the Presence of Unknown Interference \thanks{We are grateful to Giulio Grossi, Andrea Ichino, Guido Imbens, Alessandro Tarozzi, Davide Viviano, and seminar participants at the European University Institute, IRCrES, iCEBDA 2025, EuroCIM 2026, and ACIC 2026 for valuable comments and suggestions. All errors are ours.
}
\vspace{.8 cm}
}
\vspace{.8 cm}
\author{Fabrizia Mealli\thanks{Professor, European University Institute and University of Florence.}  \and Javier Viviens\thanks{Ph.D. Student, European University Institute.} } 
\date{\vspace{.8 cm} July 20,  2026 
}
\maketitle

\begin{abstract}
\noindent
% 150 words
%The stable unit treatment value (SUTVA) is a crucial assumption in the Difference-in-Differences (DiD) research design. It rules out hidden versions of treatment and any sort of interference and spillover effects across units. Even if this is a strong assumption, it has not received much attention from DiD practitioners and, in many cases, it is not even explicitly stated as an assumption, especially the no-interference assumption. In this technical note, we investigate what the DiD estimand identifies in the presence of unknown interference. We show that the DiD estimand identifies a contrast of causal effects, but it is not informative on any of these causal effects separately, without invoking further assumptions. Then, we explore different sets of assumptions under which the DiD estimand becomes informative about specific causal effects. We illustrate these results by revisiting the seminal paper on minimum wages and employment by \citet{card_minimum_1994}.

% 100 words
 The Stable Unit Treatment Value Assumption (SUTVA) is a crucial assumption in Difference-in-Differences (DiD) research designs. It rules spillover effects across units and groups. Although strong, it has received little formal attention from DiD practitioners and it is often not explicitly stated. In this paper, we show that under unknown interference, the DiD estimand identifies a contrast of causal effects but is not informative about any single effect separately. We then explore sets of assumptions under which the DiD estimand would become informative about specific causal effects. We illustrate these results by revisiting the seminal minimum wage and employment study by \citet{card_minimum_1994}.

\vspace{0.5 cm}
\noindent \textbf{JEL-Code:} C10, C21, C23.

\vspace{0.2cm}
\end{abstract}

\thispagestyle{empty}

\newpage

\setcounter{page}{1}

%%%%%%%%%%%%%%%%%%%%%%%%%%%%%%%%%%%%%%%%%%%%%%%%%%%%%%%%%%%%%%%%%%%%
%  SECTION 1: INTRODUCTION
%%%%%%%%%%%%%%%%%%%%%%%%%%%%%%%%%%%%%%%%%%%%%%%%%%%%%%%%%%%%%%%%%%%%
\section{Introduction} \label{sec:intro}

The Stable Unit Treatment Value Assumption (SUTVA) \citep{rubin_randomization_1980,imbens_causal_2015} plays a crucial role in the Difference-in-Differences (DiD) research design. This assumption rules out hidden versions of treatment and any form of interference or spillover effects across units, particularly between units belonging to different groups. Despite its importance, SUTVA has received limited attention in DiD research and is often left implicit rather than explicitly stated. Notably, several recent and influential contributions to the DiD literature \citep{de_chaisemartin_two-way_2020,callaway_difference--differences_2021,sun_estimating_2021,goodman-bacon_difference--differences_2021,wooldridge_two-way_2021} do not mention SUTVA when the existence of potential outcomes is postulated, albeit this assumption is embedded in the notation used. In a recent survey, \citet{arkhangelsky_causal_2024} argue that ``there has been little attention paid to models allowing for such interference in the recent causal panel data literature to date.''

This technical note investigates what the DiD estimand identifies in the presence of unknown interference. Although applied researchers generally recognize that interference complicates the interpretation of DiD estimates, the precise nature of this complication is often left implicit. The goal of this note is therefore not to introduce a new concern, but to provide a transparent framework that makes the identification problem explicit and to clarify exactly what the DiD estimand recovers. We demonstrate that the DiD estimand identifies a contrast of causal effects but is not informative on any of these causal effects separately. Specifically, under a modified parallel trends assumption, it identifies the difference between the total effect for treated units and the indirect effect for control units. While the researcher can test whether this difference is zero, they cannot make inference about the magnitude or the direction of either of the underlying effects.

We then propose alternative sets of assumptions under which the DiD estimand would become informative about specific causal effects. These assumptions are explicit formulations of arguments that applied researchers often invoke \emph{implicitly} when interpreting DiD estimates in settings where interference may be present. By stating these assumptions formally, we clarify the conditions under which the DiD estimand can be given a causal interpretation. By making them explicit, we could discuss possible violations and improve the intuition when these assumptions may be of concern.

Specifically, the assumptions leverage information on the trend of the potential outcome in the absence of treatment and on the nature of the interference and allow to partially identify and to disentangle the total effect on the treated and the indirect effect on the control separately. We illustrate our results by revisiting the seminal paper on minimum wages and employment by \citet{card_minimum_1994}. We discuss why the no-interference assumption embedded in SUTVA may be violated in their application, how such violations impact the interpretation of their DiD estimates, and which additional assumptions would be needed to maintain the validity of their original conclusions. We also relate our framework to some recent empirical studies \citep{bleemer_affirmative_2022,braghieri_social_2022,braakmann_expected_2024,campos_impact_2024,fenizia_organized_2024,lu_how_2024,chen_regulating_2025,truffa_undergraduate_2025}, showing how our formalization clarifies the interpretation of their DiD estimands and provides a structured foundation for arguments that are often made informally.

Recent studies have begun to explore DiD settings under interference. \cite{butts_difference--differences_2023} examines DiD in settings with spatial spillovers. The key assumption is that interference across units decays with distance and that some units are sufficiently isolated from treated ones to remain unaffected. \cite{xu_difference--differences_2025} proposes doubly robust estimators for the direct average treatment effect on the treated as well as the average spillover effects in the DiD framework from a finite population perspective. This estimation strategy relies on defining an exposure mapping, which maps treatment assigned to exposure received \citep{aronow_estimating_2017}. \cite{sun_difference--differences_2025} consider DiD under network interference. Similarly, \cite{grossi_direct_2023} and \cite{hettinger_doubly_2024} estimate causal effects in longitudinal settings under the assumption of partial interference. 

In this technical note, we remain agnostic about the form that interference takes, allowing for the possibility that all units may interfere with one another. Therefore, we do not impose any form of partial interference, specify an exposure mapping, or restrict our analysis to finite samples. Instead, we adopt the superpopulation approach \citep{imbens_causal_2015}, standard in the DiD literature \citep{arkhangelsky_causal_2024}. 

The remainder of this technical note is organized as follows. Section \ref{sec:meth} introduces the  Difference-in-Differences research design and presents the main identification results under interference. Section \ref{sec:alternatives} proposes alternative assumptions under which the \emph{total effect} on the treated and the \emph{indirect effect} on the control can be (partially) identified. Section \ref{sec:application} revisits \cite{card_minimum_1994} in light of this technical note. Section \ref{sec:conclu} concludes.

%%%%%%%%%%%%%%%%%%%%%%%%%%%%%%%%%%%%%%%%%%%%%%%%%%%%%%%%%%%%%%%%%%%%
%  SECTION 2: Methodology
%%%%%%%%%%%%%%%%%%%%%%%%%%%%%%%%%%%%%%%%%%%%%%%%%%%%%%%%%%%%%%%%%%%%
\section{Difference-in-Differences under interference} \label{sec:meth}
This section introduces the canonical DiD research design and examines what the DiD estimand identifies in the presence of interference. For simplicity, we consider the two-group, two-period case. Similar arguments can be employed for the multiple-group, multiple-period case.

\subsection{Notation and set-up}
We consider a panel data setting with $N$ units observed in two periods, $t \in \{ 0, 1\}$. Let $W_{it}$ denote a binary treatment indicator equal to 1 if unit $i$ received a treatment of interest at period $t$, and 0 otherwise. In the pre-treatment period, $t = 0$, no unit is treated ($W_{i0} = 0,\quad \forall i$). In the post-treatment period, $t = 1$, some units receive the treatment. Define $G_{i}$ as a group indicator, equal to 1 if the unit is treated in the second period and 0 otherwise. Formally, $G_{i} = 1 \iff (W_{i0},W_{i1}) = (0,1)$ and $G_{i} = 0 \iff (W_{i0},W_{i1}) = (0,0)$. 

Let $Y_{}$ denote the outcome of interest. We postulate the existence of $2^{N\times 2}$ potential outcomes for unit $i$ at time $t$, $Y_{it}(\bs{W})$, where $\bs{W}$ is the $N \times 2$ treatment assignment matrix containing the potential treatment assignment for all units in all periods. We can decompose this matrix into $N\times1$ vectors of treatment assignment, one in each period, $Y_{it}(\bs{W_{0}},\bs{W_{1}})$, where $\bs{W_{t}}$ denotes the treatment assignment vector at time $t$. In the specific setting we consider, no unit is treated in period $t = 0$, so we can write $Y_{it}(\bs{0},\bs{W_{1}})$.

\begin{assumption} No treatment anticipation. \label{as:no_anti} \\ 
    $Y_{i0}(\bs{{0}},\bs{W_{1}}) = Y_{i0}(\bs{{0}},\bs{W_{1}^{\prime}}) = Y_{i0}(\bs{{0}})  \quad \forall \bs{W_{1}}, \bs{W_{1}^{\prime}}$
\end{assumption}
Assumption \ref{as:no_anti} states that outcomes do not depend on future treatments. This assumption implies that treatments cannot affect outcomes measured before their implementation.

Therefore, we can express the potential outcomes in the pre-treatment period as a function of the contemporaneous treatment vector only, $Y_{i0}(\bs{W_{0}}) = Y_{i0}(\bs{0})$. 
Since no unit is treated at $t=0$ by design, the potential outcomes at $t=1$ can be indexed exclusively by the treatment assignments at that time, $\bs{W_{1}}$. In the two-group two-period case, under interference and the no-anticipation assumption, each unit is characterized by $2^{N}$ potential outcomes in each period.

We can summarize these $2^N$ potential outcomes by partitioning the treatment assignment vector at period $t$, $\bs{W_{t}}$, into three components: $\left(W_{it}, \bs{W_{G_{i}t}},\bs{W_{1-G_{i}t}}\right)$, where $W_{it}$ denotes the treatment of unit $i$, $\bs{W_{G_{i}t}}$ denotes the vector of treatments for units in the same group of unit $i$ and $\bs{W_{1-G_{i}t}}$ denotes the vector of treatments for units opposite group of unit $i$. This notation allows for unknown interference while remaining agnostic about its specific form, both within and across groups. 
Out of these $2^N$ potential outcomes, only two are observable in the post-treatment period, $Y_{i1}(1,\bs{1},0)$ for treated units and $Y_{i1}(0,\bs{0},\bs{1})$ for control units.
\begin{assumption} No hidden version of treatment
    \label{as:no_hid} \\
    $Y_{i0} = Y_{i0}(\bs{0})$ \\
    $Y_{i1} = Y_{i1}(1,\bs{1},\bs{0}) W_{i1} + Y_{i1}(0,\bs{0},\bs{1})(1-W_{i1})$
\end{assumption}
Assumption \ref{as:no_hid}, also known as consistency \citep{vanderweele_causal_2013}, links the observed outcome for each unit to a single well-defined potential outcome.

The Difference-in-Differences research design is typically used in observational settings where the researcher has no control over the treatment assignment. That is why in this technical note we consider assignment-conditional causal effects, where we take the treatment assignment vector as given and we do not marginalize over all the possible treatment assignments \citep{savje_average_2021}. All the expectations throughout the note are meant to be super-population expectations \citep{imbens_causal_2015}.

\subsection{Canonical DiD under the no-interference assumption}
In the canonical Difference-in-Differences research design, a maintained assumption is the no-interference assumption part of SUTVA, although it is often not explicitly stated. 
\begin{assumption} \label{as:sutva}
    No-interference
    \begin{equation*}
        Y_{it}(W_{it}, \bs{W_{G_it}},\bs{W_{1-G_it}}) =Y_{it}(W_{it}, \bs{W_{G_it}^{\prime}},\bs{W_{1-G_it}^{\prime}}) = Y_{it}(W_{it})  \quad \forall  \bs{W_{G_i,t}},\bs{W_{1-G_it}}, \bs{W_{G_it}^{\prime}},\bs{W_{1-G_it}^{\prime}}
    \end{equation*}
\end{assumption}
Assumption \ref{as:sutva} states that potential outcomes of unit $i$ do not depend on other units' treatment. It rules out both interference between and within groups. This assumption is strong and is likely to be violated in many scenarios. Under Assumption \ref{as:sutva}, it suffices to index the potential outcomes only with the unit's own treatment, which is why it is embedded in the notation used by researchers, even if it is not explicitly stated \citep{arkhangelsky_causal_2024}.

The target estimand in the canonical DiD research design is the Average Treatment effect on the Treated (ATT)
\begin{estimand}
    \label{est:att} Average Treatment effect on the Treated (ATT)
    \begin{equation*}
        \E[Y_{i1}(1) - Y_{i1}(0) \mid G_{i} = 1]
    \end{equation*}
\end{estimand}
It is clear that, without Assumption \ref{as:sutva}, this estimand is ill-defined. Estimand \ref{est:att} captures the expected effect of treating a single unit drawn from the treatment group. The DiD research design is typically applied to observational settings where the treatment is confounded with unobservable characteristics. Therefore, the identification of the ATT estimand relies on assumptions on the evolution of these unobservable confounders, given Assumptions \ref{as:no_anti} and \ref{as:sutva}. The main identification assumption in the canonical DiD framework is the Parallel Trends assumption.
\begin{assumption} \label{as:pt_sutva}
    Parallel Trends under no-interference
    \begin{equation*}
        \E[Y_{i1}(0)-Y_{i0}(0) \mid G_{i} = 1] = \E[Y_{i1}(0)-Y_{i0}(0) \mid G_{i} = 0]
    \end{equation*}
\end{assumption}
Assumption \ref{as:pt_sutva} states that, in the absence of treatment, the difference in expected outcomes between the two groups is constant over time. This Parallel Trends assumption has been the primary focus of the criticism and discussion around the DiD research design (e.g., \cite{rambachan_more_2023,roth_when_2023,huber_joint_2024,ghanem_selection_2025}). Under the parallel trends assumption, the ATT is identified by the Difference-in-Differences estimand:
\begin{estimand} Difference-in-Differences (DiD)
\begin{equation*} \label{est:did}
    DiD = \E[Y_{i1}-Y_{i0} \mid G_{i} = 1] - \E[Y_{i1}-Y_{i0} \mid G_{i} = 0].
\end{equation*}
\end{estimand}
\begin{proposition} \label{prop:can}
    Under Assumptions \ref{as:no_anti}, \ref{as:no_hid}, \ref{as:sutva} and \ref{as:pt_sutva}, the ATT (Estimand \ref{est:att}) is identified by the DiD estimand.

\begin{proof} \begin{align*}
        DiD &= \E[Y_{i1}-Y_{i0} \mid G_{i} = 1] - \E[Y_{i1}-Y_{i0} \mid G_{i} = 0] \\
        &=\E[Y_{i1}(1) \mid G_{i} = 1] - \left(\E[Y_{0}(0) \mid G_i = 1]+ \E[Y_{i1}(0)-Y_{i0}(0) \mid G_{i} = 0]\right) \\
        &=\E[Y_{i1}(1) \mid G_{i} = 1] - \E[Y_{i1}(0) \mid G_{i} = 1] = ATT,
    \end{align*}
    where Assumptions \ref{as:no_anti}, \ref{as:no_hid}, and \ref{as:sutva} are used in the second equality to  link  observed outcomes to potential outcomes, and the third equality follows from Assumption \ref{as:pt_sutva}. 
\end{proof}
\end{proposition}
Notice that the DiD estimand is not a contrast of potential outcomes on the same set of units and, therefore, it is not a causal estimand on its own. Nonetheless, it corresponds to the ATT under the set of assumptions of the canonical DiD research design (including the no-interference assumption).

\subsection{DiD under partial interference}
Next, we examine the case of partial interference, where units within the same group may interfere with one another but not with units in different groups.

\begin{assumption} \label{as:partial_int}
    Partial interference
    \begin{equation*}
        Y_{it}(W_{it}, \bs{W_{G_it}},\bs{W_{1-G_it}}) =Y_{it}(W_{it}, \bs{W_{G_it}^{}},\bs{W_{1-G_it}^{\prime}}) = Y_{it}(W_{it}, \bs{W_{G_{i}t}})  \quad \forall  \bs{W_{1-G_it}}, \bs{W_{1-G_it}^{\prime}}
    \end{equation*}
\end{assumption}
Assumption \ref{as:partial_int} rules out interference between treated and control units, similar to the SUTNVA assumption in \cite{forastiere_identification_2021}. This assumption is a weaker version of the no-interference assumption embedded in Assumption \ref{as:sutva}. When Assumption \ref{as:sutva} holds, Assumption \ref{as:partial_int} also holds. However, the opposite is not true, because even when Assumption \ref{as:partial_int} holds, there might be spillovers among the treated units. The immediate consequence of this relaxation is that the main target estimand in the DiD setting, the ATT, is no longer well-defined, since potential outcomes may now also depend on the treatment assignment of other units in the same group. 
Under partial interference, the following causal estimand is well-defined:

\begin{estimand}
    \label{est:tatt_pi} Total Average Treatment Effect on the Treated under partial interference (TATT-pi)
    \begin{equation*}
        \E[Y_{i1}(1,\bs{1}) - Y_{i1}(0,\bs{0}) \mid G_{i} = 1]
    \end{equation*}
\end{estimand}

Estimand \ref{est:tatt_pi} captures the average effect for a treated unit of treating all the units in its group.
This effect encompasses the direct treatment effect on the treated unit as well as the spillover effect from treating the other units in the group. Therefore, this estimand does not capture the effect of treating a single unit (in isolation) but rather the effect of treating the whole group. When no-interference holds, Estimand \ref{est:tatt_pi} is equivalent to the standard ATT.

Under partial interference, we also need to modify the parallel trends assumption as follows:
\begin{assumption}
    \label{as:pt_pi}
    Parallel trends under partial interference
    \begin{equation*}
        \E[Y_{i1}(0,\bs{0})-Y_{i0}(0,\bs{0}) \mid G_{i} = 1] = \E[Y_{i1}(0,\bs{0})-Y_{i0}(0,\bs{0}) \mid G_{i} = 0].
    \end{equation*}
\end{assumption}
The interpretation of this assumption remains unchanged: in the absence of treatment, the difference in expected outcomes between the two groups would remain constant over time.

\begin{proposition} \label{prop:pi}
    Under Assumptions \ref{as:no_anti}, \ref{as:no_hid}, \ref{as:partial_int} and \ref{as:pt_pi}, the TATT-pi (Estimand \ref{est:tatt_pi}) is identified by the DiD estimand.

\begin{proof} \begin{align*}
        DiD &= \E[Y_{i1}-Y_{i0} \mid G_{i} = 1] - \E[Y_{i1}-Y_{i0} \mid G_{i} = 0] \\
        &=\E[Y_{i1}(1,\bs{1}) \mid G_{i} = 1] - \left(\E[Y_{0}(0,\bs{0}) \mid G_i = 1]+ \E[Y_{i1}(0,\bs{0})-Y_{i0}(0,\bs{0}) \mid G_{i} = 0]\right) \\
        &=\E[Y_{i1}(1,\bs{1}) \mid G_{i} = 1] - \E[Y_{i1}(0,\bs{0}) \mid G_{i} = 1] = TATT-pi,
    \end{align*}
    where assumptions \ref{as:no_anti}, \ref{as:no_hid}, and \ref{as:partial_int} are used in the second equality to link observed outcomes to potential outcomes, and the third equality follows from Assumption \ref{as:pt_pi}. 
\end{proof}
\end{proposition}

Proposition \ref{prop:pi} states that the DiD estimand has a causal interpretation when units interfere within groups, provided that there is no interference between groups. However, it no longer identifies the ATT, which is not well-defined without Assumption \ref{as:sutva}, but the TATT-pi. This distinction changes the interpretation of the DiD estimand. Now, it incorporates both the average direct treatment effect and the average spillover effect on the treated. Thus, it should be interpreted as the expected effect when treating the whole group, rather than a single unit. The gap between these two interpretations becomes especially important when scaling up a policy, as changing the composition and size of the treated group may lead to different results. Note that this result is derived under Assumption \ref{as:partial_int}, which does not impose any constraint on the interference structure within the treatment group.

\subsection{DiD under unknown interference}
In this section, we consider the setting in which no restrictions are imposed on the interference structure among units. In particular, we remain agnostic about whether, and to what extent, units may interfere with one another, thereby allowing for the possibility that interference occurs both within and between groups, and potentially among all units in the population in an unrestricted manner. 

We now introduce two estimands, which are well-defined under the presence of any sort of interference. 
\begin{estimand}
    \label{est:tatt} Total Average Treatment Effect on the Treated (TATT) under unknown interference
    \begin{equation*}
        \tau^{1} = \E[Y_{i1}(1,\bs{1},\bs{0}) - Y_{i1}(0,\bs{0},\bs{0}) \mid G_{i} = 1]
    \end{equation*}
\end{estimand}
Estimand \ref{est:tatt} has a similar interpretation as \ref{est:tatt_pi}. Since we now allow for interference between treated and control units, Estimand \ref{est:tatt_pi} is not well-defined, and we must index the potential outcomes with the treatment of all units\footnote{Note that under partial interference the TATT is equivalent to the all-or-nothing effect \cite{savje_average_2021} on the treated. Under unknown interference that is no longer true, that is, Estimand \ref{est:tatt} is not equivalent to $\E[Y_{i1}(1,\bs{1},\bs{1}) - Y_{i1}(0,\bs{0},\bs{0}) \mid G_{i} = 1]$}. 

\begin{estimand}\label{est:ind}
    Average spillover effect on the control (ASC)
    \begin{equation*} 
        \tau^{0} = \mathbb{E}[Y_{i1}(0,\bs{0},\bs{1}) - Y_{i1}(0,\bs{0},\bs{0}) \mid G_{i } = 0]
    \end{equation*}
\end{estimand}
Estimand \ref{est:ind} captures the causal effect of the treatment on the non-treated units. It accounts for the average spillover effect from the treated to control units or any effect on the control units derived from the general equilibrium effects of the intervention $\bs{W_1}$. Under the no-interference assumption, this effect is zero as both potential outcomes are the same.

Next, we introduce the modified parallel trends assumption in the presence of unknown interference.
\begin{assumption}
    \label{as:pt_ui} Parallel trends under unknown interference
    \begin{equation*}
        \E[Y_{i1}(0,\bs{0},\bs{0})\mid G_{i} = 1]-\E[Y_{i1}(0,\bs{0},\bs{0}) \mid G_{i} = 0] = \E[Y_{i0}(0,\bs{0},\bs{0})\mid G_{i} = 1]-\E[Y_{i0}(0,\bs{0},\bs{0}) \mid G_{i} = 0] 
    \end{equation*}
\end{assumption}
Assumption \ref{as:pt_ui} states that, in the absence of the intervention, the difference in expected outcomes between the two groups would have remained constant over time. Therefore, the interpretation of this parallel trends assumption is equivalent to the parallel trends assumption under no-interference (Assumption \ref{as:pt_sutva}) and the parallel trends assumption under partial interference (Assumption \ref{as:pt_pi})\footnote{In this note, we consider estimands and assumptions whose analogs under SUTVA are well-defined. It is still possible to explore similar parallel trends assumption on other observed potential outcomes under unknown interference, like $Y_{it}(1,\bs{1},\bs{0})$ or $Y_{it}(0,\bs{0},\bs{1})$, under which identifications of estimands that are well defined only under interference, like the average spillover effects for the treated, could follow.}. 

\begin{proposition} \label{prop:ui}
     Under Assumptions \ref{as:no_anti}, \ref{as:no_hid}, and \ref{as:pt_ui}, the DiD estimand identifies the difference between the TATT (Estimand \ref{est:tatt}, $\tau^1$) and the ASC (Estimand \ref{est:ind}, $\tau^0$)

\begin{proof} \begin{align*}
        DiD &= \E[Y_{i1}-Y_{i0} \mid G_{i} = 1] - \E[Y_{i1}-Y_{i0} \mid G_{i} = 0] \\
        &= \E[Y_{i1}(1,\bs{1},\bs{0}) - Y_{i0}(0,\bs{0},\bs{0})\mid G_{i} = 1] - \E[Y_{i1}(0,\bs{0},\bs{1}) - Y_{i0}(0,\bs{0},\bs{0})\mid G_{i} = 0 ] \\
        &= \E[Y_{i1}(1,\bs{1},\bs{0})\mid G_i = 1] - \E[Y_{i1}(0,\bs{0},\bs{1})\mid G_i =0] - (\E[Y_{i0}(0,\bs{0},\bs{0})\mid G_{i} = 1]  - \E[Y_{i0}(0,\bs{0},\bs{0})\mid G_{i} = 0 ]) \\
         &= \E[Y_{i1}(1,\bs{1},\bs{0})\mid G_i = 1] - \E[Y_{i1}(0,\bs{0},\bs{1})\mid G_i =0] - (\E[Y_{i1}(0,\bs{0},\bs{0})\mid G_{i} = 1]  - \E[Y_{i1}(0,\bs{0},\bs{0})\mid G_{i} = 0 ]) \\
         &=\E[Y_{i1}(1,\bs{1},\bs{0}) - Y_{i1}(0,\bs{0},\bs{0})\mid G_{i} = 1] - \E[Y_{i1}(0,\bs{0},\bs{1}) - Y_{i1}(0,\bs{0},\bs{0})\mid G_{i} = 0 ] \equiv \tau^{1} - \tau^{0}
    \end{align*}
    where Assumption \ref{as:no_anti} and  \ref{as:no_hid} are used in the second equality to postulate the potential outcomes and link them to the observed outcomes, the third equality just rearranges terms, the fourth equality follows from Assumption \ref{as:pt_ui}, and the fifth equality just rearranges terms back. 
\end{proof}
\end{proposition}

 Proposition \ref{prop:ui} states that the Difference-in-Differences estimand does not identify a causal effect, but rather a difference of causal effects. It cannot be interpreted as the causal impact of the policy but rather as the difference in impacts between the treatment and the control groups.\footnote{Similar results are derived by \cite{sobel_what_2006} and \cite{vazquez-bare_identification_2023} in the context of randomized experiments for difference-in-means estimators, by \cite{xu_factorial_2025} in the context of factorial designs, and by \cite{butts_difference--differences_2023} and \cite{xu_difference--differences_2025} in the context where the interference structure is known and can be summarized by an exposure mapping. In this paper we remain agnostic about the form interference takes place and derive this general identification result in the presence of unknown interference.} Notice that estimating the DiD estimand allows for testing whether the intervention had a different average effect on the treated and control groups. Still, it does not permit testing and/or learning anything about these two different causal effects separately. If the DiD estimand is equal to 0, it could be because both $\tau^1$ and $\tau^0$ are equal to 0, or because both are equally positive or equally negative. Similarly, a positive DiD only implies that $\tau^{1} > \tau^0$. However, we can have this scenario when both effects are positive, and $\tau^{1}$ is larger than $\tau^0$, when both effects are negative, but $\tau^1$ is less negative than $\tau^0$, and when $\tau^1$ is positive and $\tau^0$ is negative. Therefore, when no-interference is violated, under a Parallel Trends assumption, the DiD estimand reveals how the impact differs across groups; however, we cannot determine the signs of the treatment effects, because for any value of the DiD estimand, there are infinite combinations of the pair of causal effects $(\tau^1,\tau^0)$ that could yield that value.

\section{Identification assumptions for TATT and ASC under unknown interference} \label{sec:alternatives}
We have demonstrated that when the no-interference assumption is violated, even if a parallel trends assumption holds, the DiD estimand lacks a causal interpretation and provides limited information on the actual treatment effect. Next, we discuss possible additional identification assumptions under which the DiD estimand could become informative about the TATT ($\tau^1$) and the ASC ($\tau^0$).

\subsection{Assumptions on the unobserved trends of $Y(0,\bs{0},\bs{0})$.}
The fundamental challenge of the DiD research design in the presence of unknown interference is that the potential outcomes in the absence of the treatment for all is not observed for any unit in the post-treatment period. However, if the researcher is willing to assume that the pre-treatment observed value of $Y_0(0,\bs{0},\bs{0})$ is informative for the unobserved post-treatment outcome, $Y_1(0,\bs{0},\bs{0})$, then they could identify $\tau^1$ and $\tau^0$ separately.

\begin{assumption} \label{as:constant}
    Constant average potential outcomes in the absence of treatment, $g\in \{0, 1\}$: $$\E[Y_{i1}(0,\bs{0},\bs{0})\mid G_{i} = g] = \E[Y_{i0}(0,\bs{0},\bs{0})\mid G_{i} = g]$$
\end{assumption}
Assumption \ref{as:constant} is the standard assumption underlying any causal interpretation of \textit{before-and-after} comparisons \citep{imbens_causal_2015}.
\begin{proposition}
    \label{prop:constant_po}
    Under Assumptions \ref{as:no_anti}, \ref{as:no_hid}, and \ref{as:constant}, $\tau^1$ and $\tau^0$ are identified as follows:
\begin{equation*}
    \tau^g = \E[Y_{i1} - Y_{i0} \mid G_{i} = g].
\end{equation*}
\begin{proof} \begin{align*}
     &  \E[Y_{i1} - Y_{i0} \mid G_{i} = 1] =  \E[Y_{i1}(1,\bs{1},\bs{0}) - Y_{i0}(0,\bs{0},\bs{0}) \mid G_i = 1] = \E[Y_{i1}(1,\bs{1},\bs{0}) - Y_{i1}(0,\bs{0},\bs{0}) \mid G_i = 1] = \tau^{1}
    \end{align*}
    where Assumption \ref{as:no_anti} and \ref{as:no_hid} are used in the first equality to link the observed outcomes to potential outcomes, and the second equality follows from Assumption \ref{as:constant}. Similar arguments can be used to identify $\tau^0$.
\end{proof}
\end{proposition}

Assumption \ref{as:constant} is strong, yet it illustrates an important insight: in the presence of interference, even if the potential outcomes $Y_{1}(0,0,0)$ were observed, we could not compare treatment and control groups to identify causal effects, as both groups are affected (potentially differently) by the treatment. A more general version of this assumption is given by:

\begin{assumption} \label{as:ctrend}
    Range of the time trend of outcome in the absence of treatment: $$ \E[Y_{it+1}(0,\bs{0},\bs{0})\mid G_{i} = g] - \E[Y_{it}(0,\bs{0},\bs{0}) \mid G_{i} = g]  \in[-k, k],$$ with $k \in \mathbb{R}_{+}$.
\end{assumption}

Assumption \ref{as:ctrend} postulates that the difference over time of the average potential outcomes lies in the range $\pm k$ (see also \cite{rambachan_more_2023} for similar assumptions in the DiD context). Under Assumption \ref{as:ctrend}, we can partially identify $\tau^1$ and $\tau^0$ as follows: 

\begin{proposition} \label{prop:range}
   Under Assumptions \ref{as:no_anti}, \ref{as:no_hid}, and  \ref{as:ctrend}, $\tau^1$ and $\tau^0$ are partially identified as follows:
\begin{equation*}
    \tau^{g}  \in \bigg[ \E[Y_{i1} - Y_{i0} \mid G_{i} = g] - k \quad , \quad  \E[Y_{i1} - Y_{i0} \mid G_{i} = g] + k \bigg]
\end{equation*}
\begin{proof} By Assumptions \ref{as:no_anti} and \ref{as:no_hid},
\begin{equation*}
    \E[Y_{i1} - Y_{i0} \mid G_{i} = 1] =  \E[Y_{i1}(1,\bs{1},\bs{0}) - Y_{i0}(0,\bs{0},\bs{0}) \mid G_i = 1], 
\end{equation*}
and from Assumption \ref{as:ctrend}, it follows that 
\begin{equation*}
    \E[Y_{i1}(0,\bs{0},\bs{0}) \mid G_i = 1] \leq  \E[Y_{i0}(0,\bs{0},\bs{0}) \mid G_i = 1] + k
\end{equation*}
and 
\begin{equation*}
     \E[Y_{i1}(0,\bs{0},\bs{0}) \mid G_i = 1] \geq \E[ Y_{i0}(0,\bs{0},\bs{0}) \mid G_i = 1] - k.
\end{equation*}
So we have that 
\begin{equation*}
    \E[Y_{i1}(1,\bs{1},\bs{0}) - Y_{i1}(0,\bs{0},\bs{0}) \mid G_i = 1] \geq \E[Y_{i1}(1,\bs{1},\bs{0}) - Y_{i0}(0,\bs{0},\bs{0}) \mid G_i = 1] - k =  \E[Y_{i1} - Y_{i0} \mid G_{i} = 1] - k
\end{equation*}
and 
\begin{equation*}
    \E[Y_{i1}(1,\bs{1},\bs{0}) - Y_{i1}(0,\bs{0},\bs{0}) \mid G_i = 1] \leq \E[Y_{i1}(1\bs{,}1,\bs{0}) - Y_{i0}(0,\bs{0},\bs{0}) \mid G_i = 1] + k =  \E[Y_{i1} - Y_{i0} \mid G_{i} = 1] + k.
\end{equation*}
Similar arguments can be used to identify $\tau^0$.
\end{proof}
\end{proposition}

If the researcher assumes a specific range of the trend on the unobserved potential outcome $Y(0,0,0)$, it would be possible to partially identify the causal effects for a given value of that trend. Again, there is no need to compare the treatment and control groups, or to assume parallel trends. This partial identification exercise can be used as a sensitivity analysis (e.g., \cite{vanderweele_sensitivity_2017,cinelli_making_2020}), as it is possible to find the value of $k$, the sensitivity parameter, such that the lower/upper bound of $\tau^g$ becomes equal to 0, that is, the slope in the linear trend required for the treatment effect to vanish.

The main takeaway from this subsection is that, in the presence of interference, the potential outcome $Y_1(0,0,0)$ is unobserved for every unit. As a result, standard parallel trends assumptions and comparisons across treated and control groups are no longer sufficient for identification. A more promising approach is to make assumptions about the time evolution of $Y_t(0,0,0)$ within each group separately, in the spirit of the time-homogeneity assumptions used in the panel data literature \citep{chernozhukov_average_2013}. This perspective is consistent with a growing literature at the intersection of causal inference and time series, which develops methods to identify causal effects without relying on a control group \citep{cerqua_causal_2024,botosaru_forecasted_2026}. Applied separately to the treated and control groups, these methods could identify $\tau^1$ and $\tau^0$ independently.

\subsection{Assumption on the causal effects $\tau^{1}$ and $\tau^{0}$}
In some scenarios, researchers' prior knowledge may justify assumptions on the causal effects of interest, $\tau^1$ and $\tau^0$. For instance, the assumption of no-interference across groups can be interpreted as imposing $\tau^0 = 0$. Depending on the type of intervention and the nature of the interference, researchers may be willing to make the following assumptions, which constrain the causal effects while remaining agnostic on how interference interplays with the outcomes.

\begin{assumption} \label{as:sign}
    Sign of the average spillover effect on the control (ASC)
    \begin{enumerate} \centering
        \item[a.] $\tau^{0} \geq 0$ 
        \item[b.] $\tau^{0} \leq 0$ 
    \end{enumerate}
\end{assumption}
\begin{proposition}\label{prop:sign}
   Under Assumptions \ref{as:no_anti}, \ref{as:no_hid}, \ref{as:pt_ui} and \ref{as:sign}, the total effect on the treated, $\tau^1$, is partially identified as follows:

   Under Assumption \ref{as:sign}a, $\tau^{1}\geq DiD$.
   
   Under Assumption \ref{as:sign}b, $\tau^{1}\leq DiD$.
   
\begin{proof} 
   From Proposition \ref{prop:ui},
   \begin{align*}
       DiD = \tau^1 - \tau^0 
   \end{align*}
   If $\tau^0 \geq 0$ (Assumption \ref{as:sign}a), then $DiD \leq \tau^1$. If $\tau^0 \leq 0$ (Assumption \ref{as:sign}b), then $DiD \geq \tau^1$.
\end{proof}
\end{proposition}

While Assumption \ref{as:sign} is strong, as it specifies the sign of the average spillover effect on the control, it still allows for heterogeneity in these spillovers. Some examples in which this assumption could be acceptable are those in which the channel of interference is well understood. For instance, clinical trials where control units also benefit from the drug through herd immunity. On the other hand, this assumption is not realistic in settings where interference arises from general equilibrium effects, making the sign of the spillover more ambiguous a priori.

Similar assumptions have been exploited by \cite{manski_how_2018}, who place bounds on the variation of treatment effects across time and space. Although the sign restriction in Assumption \ref{as:sign} yields a particularly simple interpretation of the DiD estimand as an upper or lower bound, the same logic extends to more general restrictions. For example, one could assume that $\tau^0 \geq k$, where $k$ is a sensitivity parameter chosen by the researcher. Under this assumption, it follows that $\tau^1 \leq DiD - k$. This formulation naturally lends itself to a sensitivity analysis. For instance, when the DiD estimate is positive, the researcher can ask how large $\tau^0$ must be to rule out a positive value of $\tau^1$.

\begin{assumption} \label{as:magnitude}
    Magnitude of the effects  $$\mid \tau^{1}\mid \quad \geq\quad  \mid \tau^{0}\mid $$
\end{assumption}
\begin{proposition} \label{prop:magnitude}
    Under Assumptions \ref{as:no_anti}, \ref{as:no_hid}, \ref{as:pt_ui} and \ref{as:magnitude}, the sign of $\tau^1$ coincides with the sign of the DiD estimand, $\text{sgn}(\tau^1) = \text{sgn}(DiD)$.
    
    \begin{proof}
    \begin{align*}
        \text{sgn}(DiD) = \text{sgn}(\tau^1 - \tau^0) = \text{sgn}\left(\tau^1\left(1-\frac{\tau^0}{\tau^1}\right)\right) = \text{sgn}(\tau^1)
    \end{align*}    
    where the first equality comes from Proposition \ref{prop:ui}, and the third equality from Assumption \ref{as:magnitude}, which implies that $1 - \frac{\tau^0}{\tau^1} > 0$.
    \end{proof}
\end{proposition}
Assumption \ref{as:magnitude} establishes that the \textit{total} effect on the treated units is larger in absolute value than the \textit{spillover} effect on control units. Under this assumption, the sign of $\tau^{1}$ is identified, with $\text{sgn}(\tau^1) = \text{sgn}(DiD)$, as stated by Proposition \ref{prop:magnitude}. This assumption is plausible in settings where the researcher wants to remain agnostic about the sign of the causal effects, but the nature of the interference implies that the control group cannot be more impacted than the treatment group. %The opposite assumption, $\mid \tau^{1}\mid \quad \leq\quad  \mid \tau^{0}\mid $ would imply that $\text{sgn}(\tau^0) = \text{sgn}(DiD)$.

For example, consider a study evaluating the effect of opening a new mine on a health outcome. The location of the new mine is likely confounded with many unobservable variables that also influence health, so the researchers adopt a DiD research design. They collect data on health status both before and after the mine opens. People who live close to the mine are considered treated, and people living further away are included in the control group. If the mine affects health through air pollution, it is likely that the control units are also affected by the opening, as pollutants can travel through the air, exposing all the units in the sample. As argued in this section, the DiD estimand only identifies how different these two groups are affected. However, in this case, it is reasonable to assume that, whatever the effect is, it will affect treated units more severely. From Proposition \ref{prop:magnitude}, researchers could identify the sign of $\tau^{1}$. Furthermore, if researchers also assume the sign of the spillover effect (for instance, assuming it has the same sign as the total effect for the treated), then they could interpret their DiD estimate as a lower/upper bound, following Proposition \ref{prop:sign}. 

This subsection formalizes arguments that frequently appear in the applied economics literature by making explicit the assumptions required to justify them. For example, \cite{bleemer_affirmative_2022} study California's 1998 ban on affirmative action (Proposition 209) at UC campuses and show that the policy reduced underrepresented minority (URM) students' college quality and long-run earnings, using non-URM students as a control group. They explicitly recognize that non-URM students are not a conventional control group, noting that Proposition 209 likely increased some non-URM students' admission probabilities. Consequently, they argue that their estimates capture the effect of the policy on URM students relative to its effect on non-URM students, exactly as characterized by Proposition \ref{prop:ui}. They further argue that, because there are roughly four times as many non-URM students as URM students, the policy's effect on the control group should be smaller in magnitude, maintaining the interpretation of the DiD estimate. This reasoning corresponds to Assumption \ref{as:magnitude} and Proposition \ref{prop:magnitude}. 

Another example is given by \cite{lu_how_2024}, who exploit the opening of Beijing Subway Line 15 to estimate the effect of shorter commuting times on employee productivity and compensation. They note that workers in the control group whose commuting times are unchanged may nevertheless be affected indirectly, for example because firms reward workers based on relative performance. They discuss the sign of these spillover effects and how it would determine whether the DiD estimand should be interpreted as a lower or upper bound of the \emph{causal effect}. They ultimately find evidence suggesting that the spillovers are positive, implying that the DiD estimate is a lower bound, as we argue in Proposition \ref{prop:sign}.

A similar argument appears in \cite{truffa_undergraduate_2025}, who exploit variation in the timing of university coeducation to study how increasing female undergraduate representation affects the direction of scientific research. They argue that universities treated later may already be influenced by earlier adopters, causing researchers to shift toward gender-related topics before their own institution adopts coeducation. As they note, such spillovers would bias their estimates upward. Our Proposition \ref{prop:sign} provides a formal justification for this interpretation. 

Taken together, these examples illustrate that applied researchers already rely on assumptions on $\tau^0$ and $\tau^1$ when interpreting DiD estimates. Our contribution is to make these assumptions explicit, clarify the causal conclusions that they support, and provide a common framework for discussing them without requiring a fully specified model of interference. Modeling interference remains, of course, an alternative approach when richer information on the spillover process is available.

\begin{table}[H] \centering
\begin{tabular}{|cl|}
\multicolumn{2}{c}{\textbf{Panel A: Assumptions }} \\ \hline
      \textbf{Assumption}   &  \\ \hline 
       \ref{as:no_anti} &   No treatment anticipation  \\
    \ref{as:no_hid} & No hidden version of treatment \\
    \ref{as:sutva} & No-interference  \\
    \ref{as:pt_sutva} & Parallel Trends under no-interference \\
    \ref{as:partial_int} &  Partial interference  \\ 
    \ref{as:pt_pi} & Parallel trends under partial interference  \\ 
    \ref{as:pt_ui} & Parallel trends under unknown interference \\
    \ref{as:constant} &  Constant average potential outcomes  in the absence of treatment \\
    \ref{as:ctrend} & Range of the time trend of outcome in the absence of treatment  \\ 
    \ref{as:sign} a &  $\tau^0 \geq 0$  \\  \ref{as:sign} b &  $\tau^0 \leq 0$   \\
    \ref{as:magnitude} & $\mid \tau^{1}\mid \quad \geq\quad  \mid \tau^{0}\mid $  \\
    \hline
    \end{tabular}

\vspace{0.2cm}
\begin{tabular}{|cl|}
\multicolumn{2}{c}{\textbf{Panel B:  Estimands}} \\ \hline
        \textbf{Estimand} & \\ \hline 
             $DiD$ & $\E[Y_{i1}-Y_{i0} \mid G_{i} = 1] - \E[Y_{i1}-Y_{i0} \mid G_{i} = 0]$\\
  $ATT$ &  $\E[Y_{i1}(1) - Y_{i1}(0) \mid G_{i} = 1]$ \\
    $TATT-pi$ &  $\E[Y_{i1}(1,\bs{1}) - Y_{i1}(0,\bs{0}) \mid G_{i} = 1]$ \\
     $TATT$ ($\tau^1$) & $\E[Y_{i1}(1,\bs{1},\bs{0}) - Y_{i1}(0,\bs{0},\bs{0}) \mid G_{i} = 1]$\\
$ASC$ ($\tau^0$) & $\mathbb{E}[Y_{i1}(0,\bs{0},\bs{1}) - Y_{i1}(0,\bs{0},\bs{0}) \mid G_{i } = 0]$ \\ 

    \hline
    \end{tabular}

    \vspace{0.2cm}

\begin{tabular}{|c|c|l|}
\multicolumn{3}{c}{\textbf{Panel C: Identification results}} \\ \hline
      \textbf{Proposition}& \textbf{Assumptions}&  \textbf{Identification Result} \\ \hline 
      \ref{prop:can} &   \ref{as:no_anti}, \ref{as:no_hid}, \ref{as:sutva}, \ref{as:pt_sutva} & $DiD = ATT$  \\
      \ref{prop:pi} &   \ref{as:no_anti}, \ref{as:no_hid}, \ref{as:partial_int}, \ref{as:pt_pi} &$ DiD = TATT-pi$ \\ 
      \ref{prop:ui} &   \ref{as:no_anti}, \ref{as:no_hid}, \ref{as:pt_ui} & $DiD =$ $\tau^1 - \tau^0$ \\ 
      \ref{prop:constant_po} &   \ref{as:no_anti}, \ref{as:no_hid}, \ref{as:constant}  & $ \E[Y_{i1} - Y_{i0} \mid G_{i} = g] = \tau^g$ \\
      \ref{prop:range} &   \ref{as:no_anti}, \ref{as:no_hid}, \ref{as:ctrend}  & $  \E[Y_{i1} - Y_{i0} \mid G_{i} = g] - k \leq \tau^g \leq \E[Y_{i1} - Y_{i0} \mid G_{i} = g] + k $ \\ 
      \ref{prop:sign} &   \ref{as:no_anti}, \ref{as:no_hid}, \ref{as:pt_ui}, \ref{as:sign} a & $ DiD \leq  \tau^{1}$\\
      \ref{prop:sign} &   \ref{as:no_anti}, \ref{as:no_hid}, \ref{as:pt_ui}, \ref{as:sign} b & $DiD \geq \tau^{1} $ \\
      \ref{prop:magnitude} &   \ref{as:no_anti}, \ref{as:no_hid}, \ref{as:pt_ui}, \ref{as:magnitude} & $\text{sgn}(DiD) = \text{sgn}(\tau^1) $ \\ \hline
         
    \end{tabular}
\caption{Summary of the paper. Panel A presents all the assumptions, Panel B all the estimands, and Panel C all the identification results discussed in this paper. }
\vspace{-0.2cm}

\label{tab:outcomes}
\end{table}

%%%%%%%%%%%%%%%%%%%%%%%%%%%%%%%%%%%%%%%%%%%%%%%%%%%%%%%%%%%%%%%%%%%%
%  SECTION 3: Application
%%%%%%%%%%%%%%%%%%%%%%%%%%%%%%%%%%%%%%%%%%%%%%%%%%%%%%%%%%%%%%%%%%%%
\section{Application: revisiting \cite{card_minimum_1994}} \label{sec:application}
In this section, we revisit the seminal paper by \cite{card_minimum_1994} to illustrate our results. In the year 1992, the state of New Jersey raised the minimum wage from \$4.25 to \$5.05. \cite{card_minimum_1994} investigate the effect of this minimum wage increase on employment. To do so, \cite{card_minimum_1994} interviewed a sample of fast food restaurants in New Jersey and eastern Pennsylvania, where the minimum wage remained constant, right before the implementation of the raise and 7-8 months after. Then, they estimate the effect on full-time equivalent (FTE) workers using DiD. Table \ref{tab:CardKrueger} summarizes their findings.
\begin{table}[H]
    \centering
    \begin{tabular}{c|c|c|c}
        \textbf{State} & \textbf{Pre} & \textbf{Post} & \textbf{Difference} \\ \hline 
        New Jersey  & 20.44 & 21.03 & 0.59\\
        Pennsylvania & 23.33 & 21.17 & -2.16  \\ \hline
        Difference & -2.89 & -0.14 & 2.75
    \end{tabular}
    \caption{FTE workers in New Jersey and Pennsylvania before and after New Jersey's increase in minimum wage, from Table 3 in \cite{card_minimum_1994}.}
    \label{tab:CardKrueger}
\end{table}

Table \ref{tab:CardKrueger} reports the average number of FTE workers employed in each state. The average employment in New Jersey increased slightly after the minimum wage hike, while employment in Pennsylvania decreased. This yields a DiD estimate equal to 2.75 FTE employees. Under the parallel trends assumption and no-interference, we could conclude that the increase in minimum wage increased employment in New Jersey's restaurants by 2.75 FTE workers on average, as argued by the authors who claim ``we find that the increase in the minimum wage increased employment'' \citep[p.~792]{card_minimum_1994}. 

However, a substantial body of literature in economics demonstrates that increases in minimum wages generate spillover effects (e.g., \cite{grossman_impact_1983,cengiz_effect_2019,caires_internal_2024}). Part of this literature focuses on spatial spillovers and demonstrates that the increase of minimum wages may also affect bordering regions in various ways (e.g., \cite{kuehn_spillover_2016,shirley_response_2018,mckinnish_cross-state_2017,jardim_minimum-wage_2022,jha_whats_2024}), with commuting being one of the main drivers of spillovers. In the presence of general equilibrium effects, it is reasonable to think that the increase in minimum wage in New Jersey also affected fast food restaurants in eastern Pennsylvania, thereby violating the no-interference assumption. The spillovers could arise from multiple channels and operate in different directions. 
For instance, in the context of cross-border commuting, some Pennsylvania workers may have sought employment in New Jersey to benefit from higher wages, reducing labor supply in Pennsylvania and exerting downward pressure on restaurant employment there. This mechanism could explain the observed decline in Pennsylvania employment, which drives the positive treatment effect for New Jersey. 

If interference is present, it is no longer possible to claim that the increase in the minimum wage led to an increase in employment. Instead, conditional on the parallel trends assumption (Assumption \ref{as:pt_ui}) holding, we can only determine the differential effect. That is, we could conclude that the \textit{total} average effect of the raise in minimum wage on employment in New Jersey was 2.75 FTE employees larger than the average \textit{spillover} effect in Pennsylvania. However, we cannot determine the effect on employment in any of the states. In fact, it is possible that the rise in minimum wage decreased employment in both states, with a larger decrease in Pennsylvania, resulting in a positive difference but contradicting the original claim in \cite{card_minimum_1994}.

In light of Section \ref{sec:alternatives}, there are alternative assumptions under which the conclusions in \cite{card_minimum_1994} remain valid in the presence of unknown interference. One such assumption is that, regardless of the sign of the average spillover effect in Pennsylvania, this effect is smaller in magnitude than the total effect in New Jersey, that is, Assumption \ref{as:magnitude}. Under this assumption, the DiD estimate identifies the sign of the total average effect for New Jersey's restaurants, allowing us to conclude that the minimum wage increase did raise employment. 

Another possibility is to assume that the minimum wage increase in New Jersey led to an increase in employment in Pennsylvania, that is, Assumption \ref{as:sign}a. Under this assumption, not only does the original conclusion remain valid, but the estimated increase of 2.75 FTE workers can be interpreted as the lower bound of the true causal effect for New Jersey.

Finally, we can impose assumptions on the trends of the expected outcomes in the absence of treatment, as in Assumptions \ref{as:constant} and \ref{as:ctrend}. If the average employment had remained constant in both states without the minimum wage increase, we could conclude that the policy had a positive but limited effect on New Jersey (0.59 FTE workers), and a large negative effect on the neighboring region of Pennsylvania (-2.16 FTE workers). To rule out these effects, we would need to assume that, in the absence of treatment, New Jersey was in a slight upward trend, and Pennsylvania in a sharp downward one, with employment falling by roughly 10\% in less than a year.

There are several recent examples published in high-impact journals that do not explicitly invoke SUTVA or the no-interference assumption, but nonetheless discuss potential complications arising from spillover effects. These papers challenge the interpretation of the DiD estimates, characterizing them as `general equilibrium' effects that comprise both indirect and direct effects on the treated \citep{braghieri_social_2022,campos_impact_2024}, in line with our Estimand \ref{est:tatt_pi}; bound the relative importance of the ASC ($\tau^0$) in order to maintain the conclusions derived from the DiD estimates \citep{bleemer_affirmative_2022}, in line with our results in Proposition \ref{prop:magnitude}; argue about the direction of the bias coming from the spillovers and how the DiD estimate can be interpreted as a lower bound \citep{lu_how_2024,truffa_undergraduate_2025}, in line with our results in Proposition \ref{prop:sign}; or try to disentangle the $TATT$ ($\tau^1$) and ASC ($\tau^0$) by imputing the missing potential outcome $Y_{i1}(0, \bs{0},\bs{0})$ either by finding a control group unaffected by the spillovers \citep{fenizia_organized_2024} or by defining an exposure mapping to model interference \citep{braakmann_expected_2024,chen_regulating_2025}. By postulating the complete set of potential outcomes and defining causal estimands in the presence of spillovers, this technical note aims to clarify and add structure to such discussions.

%%%%%%%%%%%%%%%%%%%%%%%%%%%%%%%%%%%%%%%%%%%%%%%%%%%%%%%%%%%%%%%%%%%%
%  SECTION 4: Conclusions
%%%%%%%%%%%%%%%%%%%%%%%%%%%%%%%%%%%%%%%%%%%%%%%%%%%%%%%%%%%%%%%%%%%%
\section{Concluding remarks} \label{sec:conclu}
In this note we have shown that, in the presence of unknown interference, the canonical DiD research design only identifies a contrast of causal effects, but it is not informative on any of them separately. We have provided possible assumptions under which researchers could (partially) identify policy-relevant causal effects in such settings. Throughout the paper, we have remained agnostic about the form interference takes, allowing for arbitrary patterns of spillovers. An alternative approach is to directly model interference, as a recent and growing literature has begun to do \citep{butts_difference--differences_2023,grossi_direct_2023,hettinger_doubly_2024,sun_difference--differences_2025,xu_difference--differences_2025}.

We have considered only the simplest case of two groups and two periods. However, in most applications, researchers have access to multiple periods of data \citep{roth_whats_2023}, and the intervention of interest is implemented in a staggered fashion. Our results still apply in these settings. Nevertheless, the availability of multiple periods and groups creates opportunities to explore alternative identification strategies that exploit knowledge about carryover effects and the intensity of spillovers.

\bibliography{references} 
\bibliographystyle{apalike} 
%===================================================================
%===================================================================
%APPENDIX
%===================================================================
%===================================================================
\newpage

\appendix

\renewcommand{\theequation}{A\arabic{equation}}
\renewcommand{\thefootnote}{A\arabic{footnote}}
\renewcommand{\thetable}{A\arabic{table}}
\renewcommand{\thefigure}{A\arabic{figure}}
\renewcommand{\thelemma}{A\arabic{lemma}}
\renewcommand{\thecorollary}{A\arabic{corollary}}
\setcounter{equation}{0}
\setcounter{footnote}{0}
\setcounter{table}{0}
\setcounter{figure}{0}
\setcounter{lemma}{0}
\linespread{1}

\newpage
\vspace{-0.4 cm}

\end{document}